\documentclass[parallelisme]{compas2020}
\usepackage[pagebackref=true]{hyperref}
\usepackage[utf8]{inputenc}
\usepackage{longtable}
\usepackage{float}
\usepackage{caption}
\newfloat{Code}{htbp}{loa}

\toappear{1} 

\usepackage[dvipsnames]{xcolor}

\usepackage{courier}

\usepackage{listings}
\definecolor{cppColorBackground}{rgb}{1.,1.,1.}
\definecolor{cppColorComment}{rgb}{0.0,0.47,.8}
\definecolor{cppColorLine}{rgb}{0.6,0.6,0.6}
\definecolor{cppColorString}{rgb}{0,0.501,145}
\definecolor{cppColorKey}{rgb}{0.8,0.5,0}
\definecolor{cppColorDigit}{rgb}{0,0,0.5}
\usepackage{pgfplots}
\pgfplotsset{compat=newest}
\usetikzlibrary{patterns}

\usepackage{subcaption}
\definecolor{color_mkl}{rgb}{0.3, 0.3, 0.3}
\definecolor{color_csr5}{rgb}{0.7, 0.75, 0.71}

\definecolor{color_b_1_8_wt}{rgb}{0.0, 0.3, 0.8}
\definecolor{color_b_2_4_wt}{rgb}{0.8, 0.0, 0.0}
\definecolor{color_b_2_4}{rgb}{0.75, 0.0, 0.2}
\definecolor{color_b_2_8}{rgb}{0.5, 0.0, 0.1}
\definecolor{color_b_4_4}{rgb}{0.0, 0.34, 0.25}
\definecolor{color_b_4_8}{rgb}{0.4, 0.7, 0.0}
\definecolor{color_b_8_4}{rgb}{0.53, 0.47, 0.76}

\usepackage[compact]{titlesec}
\titlespacing{\section}{0pt}{*1}{*1}
 \titlespacing{\subsection}{0pt}{*0.5}{*0.5}
\titlespacing{\subsubsection}{0pt}{*0.5}{*0}
\titlespacing{\paragraph}{0pt}{*0.5}{*0}

\begin{document}

\title{Automatic task-based parallelization of C++ applications by source-to-source transformations}
\shorttitle{Automatic task-based parallelization}
\author{Garip Kusoglu, Berenger Bramas, Stephane Genaud}
\address{CAMUS Inria Nancy -- Grand Est, Strasbourg University, ICPS ICube,\\
	300 bd Sébastien Brant,	67412 Illkirch - France\\
	Garip.Kusoglu@inria.fr}
\date{\today}
\maketitle
\begin{abstract}
    Currently, multi/many-core CPUs are considered standard in most types of computers including, mobile phones, PCs or supercomputers.
    However, the parallelization of applications as well as refactoring/design of applications for efficient hardware usage remains restricted to experts who have advanced technical knowledge and who can invest time tuning their software.
    In this context, the compilation community has proposed different methods for automatic parallelization, but their focus is traditionally on loops and nested loops with the support of polyhedral techniques.
    In this study, we propose a new approach to transform sequential C++ source code into a task-based parallel one by inserting annotations.
    We explain the different mechanisms we used to create tasks at each function/method call, and how we can limit the number of tasks.
    Our method can be implemented on top of the OpenMP 4.0 standard. It is compiler-independent and can rely on external well-optimized OpenMP libraries.
    Finally, we provide preliminary performance results that illustrate the potential of our method.
	\MotsCles{Task-based, Compilation, Parallelization, Object-oriented.}
\end{abstract}
\section{Introduction}
High-performance computing (HPC) is crucial for making advances and discoveries in numerous domains~\cite{EXDCI}.
While computers and supercomputers are becoming more powerful, they are also becoming increasingly more complex and heterogeneous.
The efficient utilization and programmability of such systems are ongoing research topics and remain reserved for experts who master all the technological layers and who can fine-tune their target applications to specific hardware.
Indeed, using CPUs or accelerators at an efficiency close to their theoretical peak performance is a tedious challenge especially for many-core CPUs like the Intel KNL or the Sunway SW26010~\cite{10.1145/3205289.3205313}, that make up the Sunway TaihuLight cluster, the fastest supercomputer in the world from 2016 to 2018.

This complexity has been defined as the programmability wall~\cite{10.1007/978-3-540-76837-1_3}.
To overcome this challenge and provide optimized applications to non-experts, the research community has covered multiple facets to provide methods and tools to benefit from the power of modern hardware.
On the one hand, the HPC community has proposed several parallelization paradigms, such as the task-based methods.
This paradigm has gained popularity because it allows parallelizing with an abstraction of the hardware by delegating task distribution and load balancing to dynamic schedulers.
It is a convenient solution that works well on both homogeneous and heterogeneous computing nodes.
It has proven its potential on numerous computational applications~\cite{doi10.1137/130915662,doi10.1002/cpe.3723,COUTEYENCARPAYE2018439,8053812,cs5797,COUTEYENCARPAYE2018439,7912335,qrmumps,myllykoski2019introduction}.
However, this method requires a high degree of expertise and significant programming efforts.

On the other hand, the compilation community has been creating automatic parallelization and optimization methods focusing mainly on loops with the support of the polyhedral mathematical foundation~\cite{clauss1998parametric}.
In practice, these methods work well if the time-consuming parts of code are isolated and centralized in loops that can be parallelized with a fork-join strategy.
However, many applications are not adapted because they use abstraction mechanisms, have small loops, use indirection or have loops that cannot be analyzed statically.
Therefore, its impact on real applications and the adoption by non-experts remain limited.

In this context, we propose a method that bridges the gap between automatic parallelization and task-based method.
Our method consists in transforming a C++ source code by inserting annotations to obtain a parallel application, as an HPC expert developer would do.

The contributions of the current study are:
\begin{itemize}
	\item introduce our compiler APAC and present the mechanisms to apply the task-and\\-dependency method in automatic parallelization for C++ ;
	\item describe different strategies to generate the OpenMP source code to control the number of tasks ;
	\item provide a performance study that illustrates the benefit of our method.
\end{itemize}

The paper is organized as follows.
In Section~\ref{sec:background}, we provide the background concerning the task-based method and a state of the art on automatic parallelization.
Then, in Section~\ref{sec:apac}, we describe our method and the different mechanisms in action.
Section~\ref{sec:results} includes the performance study.
Finally, we conclude in Section~\ref{sec:conclusion} and provide an extensive discussion in Section~\ref{sec:discussion}.
\section{Background}
\label{sec:background}
\subsection{Task-based programming (sequential task flow)}
The HPC community has proposed and promoted the task-based method for parallelizing scientific applications.
With this method, an application is transformed into a graph where each node is a task and each edge a dependency.
Several runtime systems have been conceived on top of this paradigm~\cite{danalis2014ptg,perez2008dependency,gautier2013xkaapi,bauer2012legion,tillenius2015superglue,augonnet2011starpu,10.7717/peerj-cs.183}, each with its features and interface~\cite{10.1007/s11227-018-2238-4}.
Underneath, they implement the paradigm using a task-graph (also called task-dependency graph) where the nodes of the graph represent operations while the edges represent the dependencies between them.
The graph materializes either within the application itself, e.g. the application uses a graph data structure to model its logic, or it materializes implicitly at runtime within a task runtime system as it is the case with the sequential task flow (STF) model.
In the STF model, a sequential code is annotated or split into tasks that are submitted to the runtime system by a single thread.
Together with the tasks, the thread provides the parameters and the type of data accesses - "read" or "(read-)write" - allowing the runtime system to generate the dependencies and to ensure that the parallel execution remains equivalent to the sequential one.

\subsection{OpenMP}
OpenMP (Open Multi-Processing)~\cite{chapman2008using} is an API that was originally dedicated to the programming of shared memory multiprocessing and which is now extended to heterogeneous computing.
It is designed for C/C++ and Fortran and is supported by most compilers, such as GNU (gcc/g++), Clang, Intel compilers, etc.
To use OpenMP, one simply needs to annotate a source code using a \emph{pragma}.
Its portability and simplicity make it a de facto standard in the development of HPC scientific applications.
For legacy reasons, it is expected that the source code would still be valid even if the compiler does not support OpenMP and omits the \emph{pragma}.
OpenMP supports the task-based method since version 4~\cite{openmp4} by letting the programmer indicate the data accesses for each task.
OpenMP supports two synchronization mechanisms to wait for the completion of tasks.
The first approach is to declare a \emph{taskgroup} scope.
A thread generates some tasks inside the \emph{taskgroup} and will leave the \emph{taskgroup} scope only once all the inserted tasks and their descendants will have completed their execution.
The second approach is the \emph{taskwait} statement that a thread can pass only when all the tasks it has inserted (but not necessarily their descendants) are terminated.

\subsection{Automatic parallelization related work}
Automatic parallelization has been investigated since the '90s to help programmers using multicore CPUs~\cite{Theory1993}.
State-of-the-art systems mostly focus on the automatic parallelization of affine loops~\cite{10.5555/1899290.1899294,ramoncorteshal-01936351,10.1007/978-0-387-72258-0_16} by considering that the workload is concentrated in loop-blocks on native data types.
For instance, the AutoPar-Clava compiler~\cite{AutoPar-Clava2018Jan,AutoPar-Clava2019Dec} uses OpenMP annotations to parallelize C code.
Meanwhile, automatic task-based parallelization has also been investigated but with less investment.
The OoOJava compiler~\cite{OoOJava2010} offers a semi-automatic parallelization process.
When using OoOJava, it is required that programmers add annotations in their source code to indicate which portions of the code should be considered as tasks.
The compiler is in charge of finding dependencies and ensuring the sequential coherency, which relieves the programmers from a significant effort.
Similarly, the @PT compiler~\cite{@PT2017} allows parallelizing an application by inserting few annotations inside JAVA codes.
Both OoOJava and @PT appear efficient and provide a way for the programmer to give information that the compiler could not know otherwise.
However, they still require programmers to modify their code to make it parallel.

The JPar compiler~\cite{JPar2016Apr} is very similar to what we want to achieve.
This project consists of a Java compiler  that analyzes the AST to find the data
accesses and to infer task boundaries.  It enables only those tasks that contain
a sufficient number  of instructions in their bodies.  However,  the compiler is
designed  for  JAVA, which  is  not  among  the  languages privileged  by  the
high-performance community. It  also has many differences with  the C++ language
and those, such as the garbage collector, impact the mechanisms used for gaining
automatic parallelization.  This compiler uses the "future" concept to model the
dependencies whereas we want to use the sequential task-based programming model.

Some attempts have also been made to use a high-level language to automatically generate task-based C++ code, such as NT$^2$ (Numerical Template Toolbox) a specific DSL similar to Matlab's language~\cite{NTT2016}.
However, the specificity of the language and its lack of standardization limit its adoption by end-users and requires complete re-writing of existing software.
\section{Source-to-source transformations for automatic task-based parallelization}
\label{sec:apac}

\subsection{Principles}
APAC is a C++ source-to-source compiler based on clang-tools v8 and the LibTooling library from the Clang subproject of the LLVM framework.
It takes as input standard C++ source code and generates C++ source code with annotations to enable task-based parallelism.
The LibTooling library makes it convenient to apply source code modifications and transformations.
Therefore, in the current implementation, we have decided to avoid working on the Clang AST.
Instead, we use LibTooling’s rewriter feature to commit changes directly to the source code buffer.
\subsection{Task creation and data access detection}

APAC applies  two main transformations  to the  input code.
It  encapsulates the body of each function into an OpenMP \emph{taskgroup}, and each method/function call into an OpenMP \emph{task} statement.
However, the calls to methods/functions from the standard  library are not encapsulated into tasks, and APAC avoid adding a \emph{taskgroup} when a function does not create any task.
With this approach, most calls  become asynchronous but a function will return only when all the  tasks it has  created (and their sub-tasks) have completed.\\

When creating a task, the  compiler makes explicit the data-dependencies through
the \texttt{depend}  clause. First,  it qualifies the  parameters passed  in the
call either  as \emph{read}  or \emph{write} variables.   For that  purpose, the
following rule is  applied: if the parameter type (callee  side) is \emph{const}
or passed by value, then the depend is \emph{in}.  Otherwise, the parameter type
is a non-\emph{const}  reference/pointer and the depend  clause is \emph{inout}.
The argument type (caller side) does not matter.

We provide in Code~\ref{code:TransformedBase} an example of parallelization with
APAC.  At  the beginning  of the  \emph{main} function,  we create  the parallel
section at  line~\ref{codex:TransformedBase:parallel}.  Then, a task  is created
at  line~\ref{codex:TransformedBase:task1}  that  includes   the  type  of  data
access.

Second, if the  call returns a result  assigned to a variable,  this variable is
also added to the \emph{inout} list. A particular case is when this statement is
used to declare  and initialize at the  same time the variable  getting the call
result.  In that  case we need to  split the statement into  its declaration and
initialization.  If the declaration qualifies the variable as \emph{const}, this
property     must      be     removed.       This     is      illustrated     in
Code~\ref{code:OriginalAssignment}      which      is      transformed      into
Code~\ref{code:TransformedAssignment}.


\subsection{Dependency expression based on previous tasks' results}
A function/method call may pass an element of an array as an argument where the index of the element is the output of a previous task.
For instance, if a task read-write \emph{i} and the next task accesses \emph{a[i]}, we must make sure that when the \emph{depend} clause is evaluated the value of \emph{i} is known.
Therefore, in this situation, we put a \emph{taskwait} before the second task, which ensures a correct dependency evaluation at the cost of constraining the degree of parallelism.

\subsection{Scope local declaration of variables}
The sequential code  may contain declarations of variables inside  a scope or in
the  body  of  an  \emph{if/else},  \emph{switch},  \emph{while}  or  \emph{for}
statements.  These  variables are destroyed at  the end of the  scope where they
are declared.  If  such a declaration is simply an  alias (pointer or reference)
to  a variable  declared outside  the  scope, nothing  has  to be  done and  the
variable  will be  passed as  \emph{firstprivate} to  the task.   Otherwise, the
compiler creates  a variable  in the heap  and uses a  reference on  the pointed
element inside the task.   The allocated element is deleted by  an extra task at
the end of the  corresponding scope.  By doing so, we  ensure that variables are
alive in the tasks  and only destroyed when no other task wants  to use it.  For
example, see line~\ref{codex:newptr} of Code~\ref{code:TransformedIfDecl}, which
was originally Code~\ref{code:OriginalIfDecl}.
\subsection{Synchronizations to enforce coherency}
We currently transform only the calls into tasks.
Therefore, the use of variables directly inside the function, between calls, are not protected by the task mechanism.
This is why accessing variables requires enforcing access coherency.
For instance, if a variable is accessed by a task and is also directly accessed later in the same \emph{taskgroup}, we need to put a \emph{taskwait} if both accesses cannot be done concurrently (which happens when both accesses are in \emph{write}, or one in \emph{read} and the other in \emph{write}).
An example of this situation is given in Code~\ref{code:TransformedSynchronization}.
\subsection{Management of returned value after task-encapsulation}
A function/method can have multiple \emph{return} statements in its body.  These
\emph{return} statements cannot  simply be put in  the tasks as we  have to make
sure that the correct  value is returned when we transform  the source code.  To
address  this  problem,  we  put  a single  \emph{return}  statement  after  the
\emph{taskgroup} and a goto-label just  before closing the \emph{taskgroups}, as
shown in  Code~\ref{code:TransformedReturn}.  Then, each  original \emph{return}
is transformed  into a \emph{taskwait}, followed  by the assignment of  the only
variable that will be  returned, and a jump to the goto-label at  the end of the
\emph{taskgroups}.
\subsection{Limiting the number of tasks}
The files generated by APAC will have \emph{taskgroup} inside each function/method and tasks at each call.
However, each task implies an overhead and thus the number of tasks should be controlled.
Ideally, the granularity should be taken into account to avoid creating tiny tasks, but we do not support this feature currently.
To control the number of tasks we use two different strategies that will set a boolean variable at the beginning of each \emph{taskgroup}.
This boolean is passed to each task to decide if it is inserted and made parallel or if it is executed directly by the thread that manages \emph{taskgroups}.
We must use the same boolean for all the tasks in a \emph{taskgroup} because it will be undefined behavior to have only some of the tasks in parallel and some other computed by the thread that manages the \emph{taskgroups}.
\subsubsection{Maximum number of existing tasks}
This strategy sets the activation boolean to true if the number of existing tasks is lower than a given number at the creation of the \emph{taskgroups}.
We have to maintain a global counter, which is incremented before each task (in the \emph{taskgroup} and just before the task statement) and decremented at the end of each task (inside the task).
If the activation boolean is set to false the global counter remains unchanged in the current \emph{taskgroups}.
\subsubsection{Maximum parallel depth}
This strategy allows us to control the call level above which the parallelism becomes disabled.
The main function is considered as level zero and the activation boolean is set to true if the current depth is lower than a given number.
To implement this mechanism, we use a thread private counter that will allow transferring the depth between \emph{taskgroups}, function calls, and tasks.
More precisely, at the beginning of the \emph{taskgroup}, the thread in charge copies its private counter into a local variable.
Then, it passes this local variable as \emph{firstprivate} to the tasks of the \emph{taskgroups}.
Finally, when a thread starts a task, it copies the \emph{firstprivate} counter + 1 into its private one and uses it in the \emph{taskgroup} it may create in the current task.
We use a thread private counter such that each branch of the tree has its own counter and allows each thread to know its current depth. 
\section{Results}
\label{sec:results}
\subsection{Configuration}
\paragraph{Software and backend}
APAC can use several runtime systems as backend.
In the current study, we focus on OpenMP that will be in charge of managing the tasks/dependencies.
It is important to understand that OpenMP is an API that can be implemented in many ways and thus performances are not portable.
For instance, because we declare several layers of \emph{taskgroups}  and tasks, the libgomp used by GNU compilers constraints the parallel execution as it appears that only one thread will be involved executing tasks from a nested \emph{taskgroups}.
This is why it is mandatory to use Clang++-8 currently and its libomp OpenMP implementation to have parallelism.
During the executions, we bind the threads using \emph{OMP\_PROC\_BIND=true}.
\paragraph{Hardware}
We perform the tests on two different configurations:
\begin{itemize}
	\item PC: Intel i7-8550U CPU at 1.80GHz with caches of sizes 32/256/8192 KB, 16 GB of RAM and 4 cores.
	\item WS: Intel Xeon Gold 6126 CPU at 2.60GHz with caches of sizes 32/1024/19712 KB, 187 GB of RAM and 20 cores.
\end{itemize}
\subsection{Test-cases}
We assess our method on two test cases~\footnote{The original and the transformed source code are available at \url{https://mybox.inria.fr/d/8fdb415c6583465eb72a/}. It is not needed to create an account but a valid email will be required.}:
\begin{itemize}
	\item Quicksort: this application sorts an array of 100 million integers.
	      The values are randomly generated and the application ends with a test to ensure the array is sorted.
	      We limit the activation of the tasks after a depth of 5.
	\item Molecular-Dyn: this application computes interactions between 10.000 particles distributed in a grid of dimension $5^3$.
	      The simulation executes 100 iterations, where one iteration consists of the computation of interactions between closed neighbors and the displacement of the particles.
\end{itemize}

The execution time is the median of 10 executions measured with the Linux \emph{time} command.
Therefore, it includes all the overhead of the parallelization system and the portions of code that have not been parallelized.
\subsection{Results}
\begin{figure}[ht!]
	\centering
	\begin{subfigure}[t]{0.24\textwidth}
		\centering
		\includegraphics[page=1, width=\textwidth, height=.20\textheight, keepaspectratio]{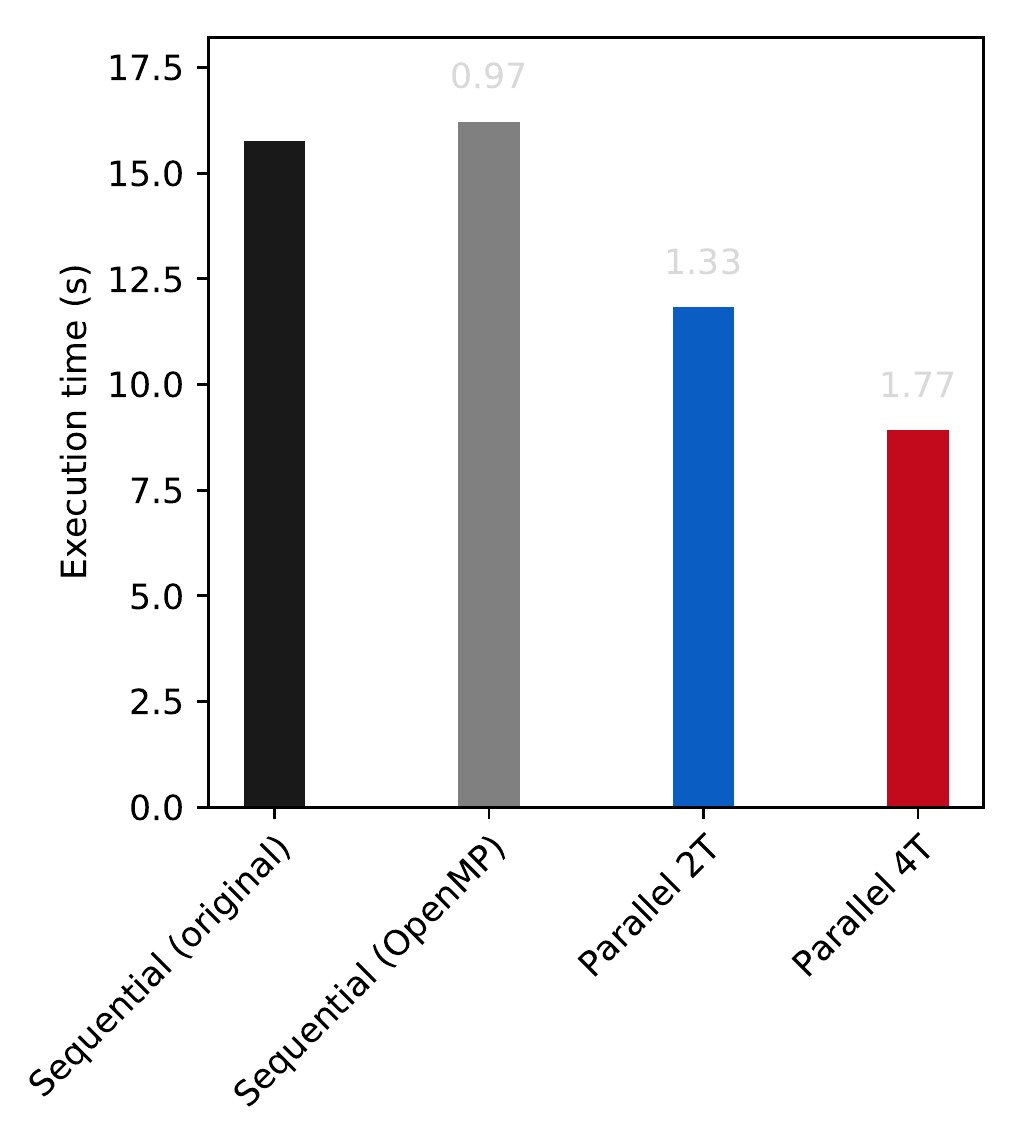}
		\caption{Quicksort PC}
		\label{fig:res-quicksort-pc}
	\end{subfigure}%
	\begin{subfigure}[t]{0.24\textwidth}
		\centering
		\includegraphics[page=1, width=\textwidth, height=.20\textheight, keepaspectratio]{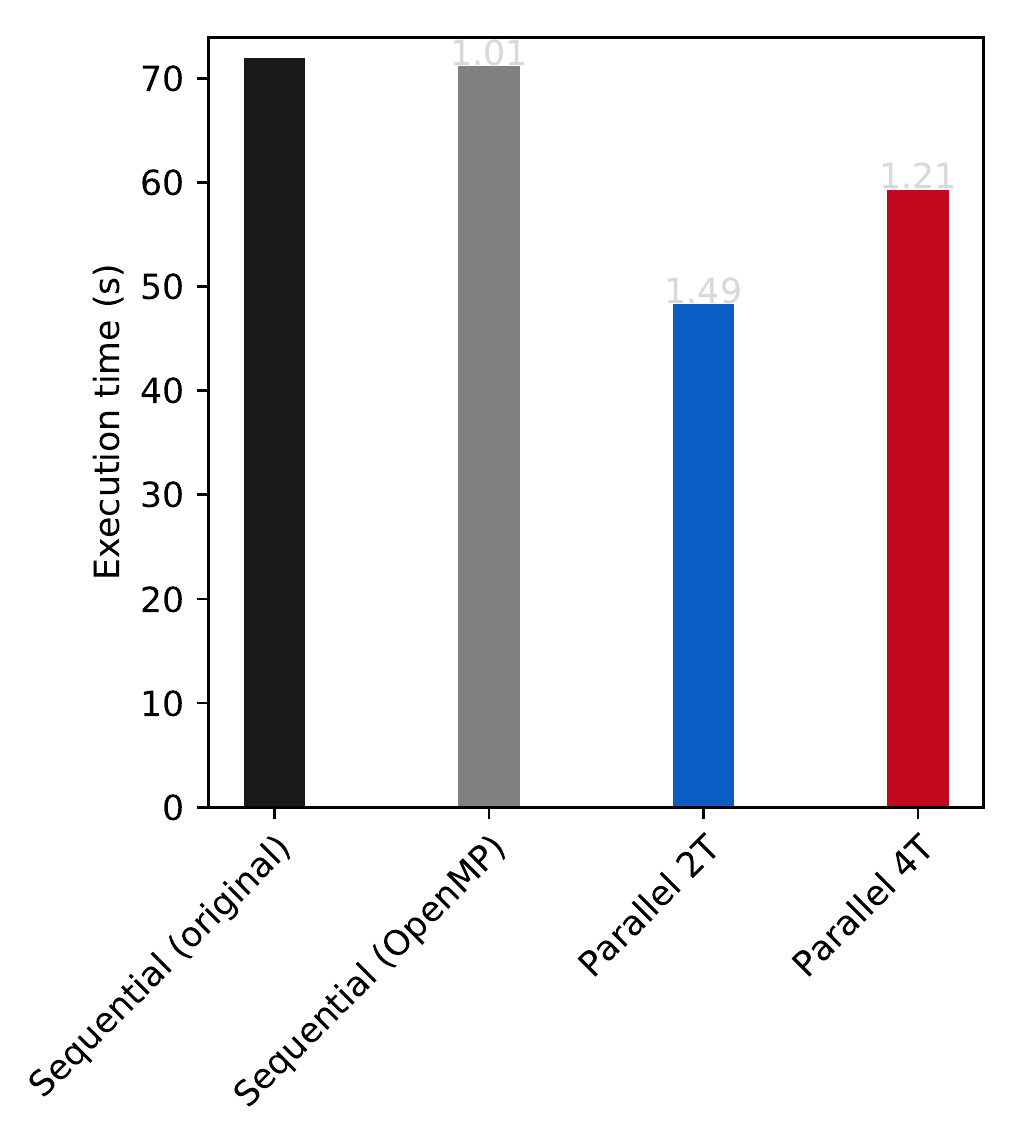}
		\caption{Molecular-Dyn PC}
		\label{fig:res-molecular-dyn-pc}
	\end{subfigure}%
    \hfill
	\begin{subfigure}[t]{0.24\textwidth}
		\centering
		\includegraphics[page=1, width=\textwidth, height=.20\textheight, keepaspectratio]{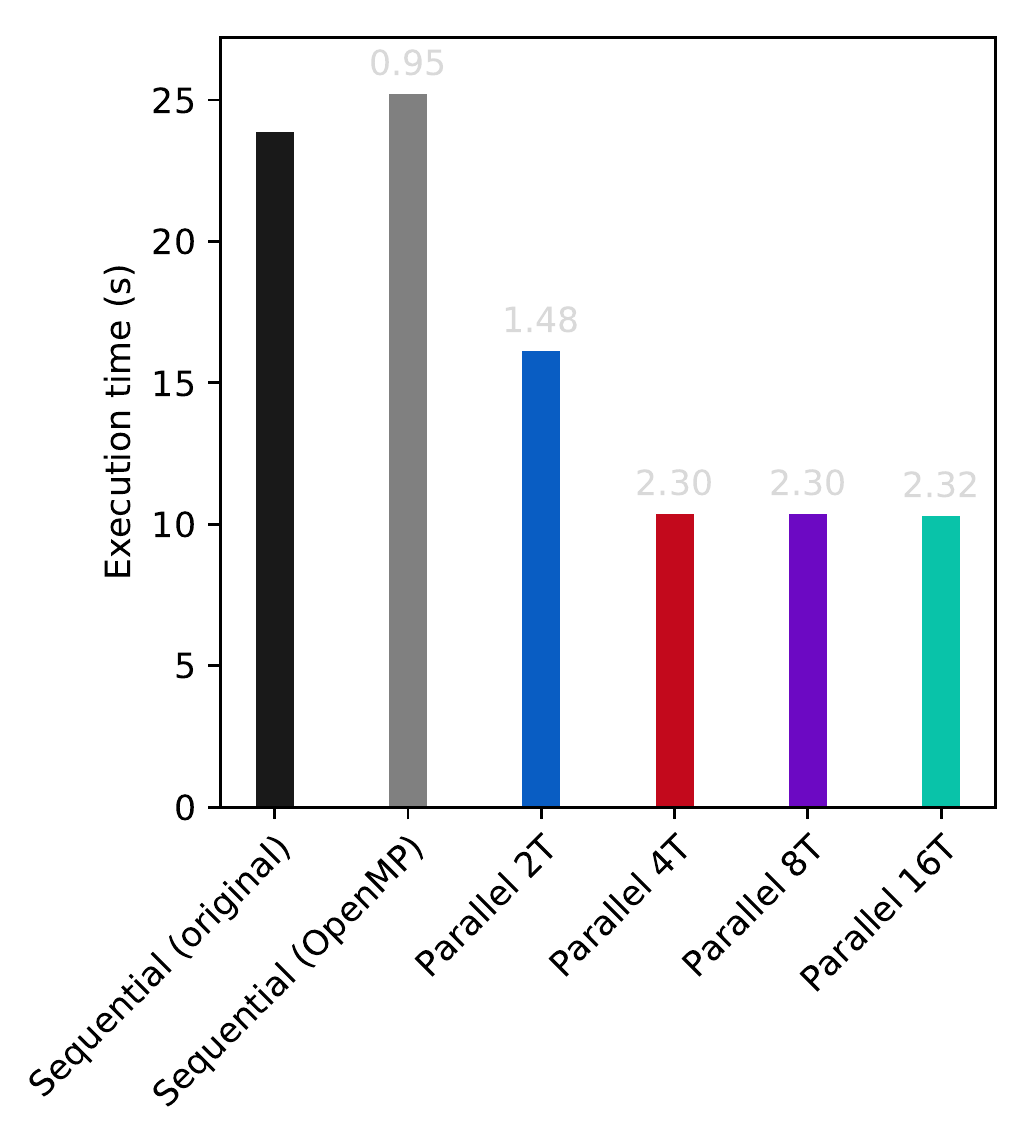}
		\caption{Quicksort WS}
		\label{fig:res-quicksort-ws}
	\end{subfigure}%
	\begin{subfigure}[t]{0.24\textwidth}
		\centering
		\includegraphics[page=1, width=\textwidth, height=.20\textheight, keepaspectratio]{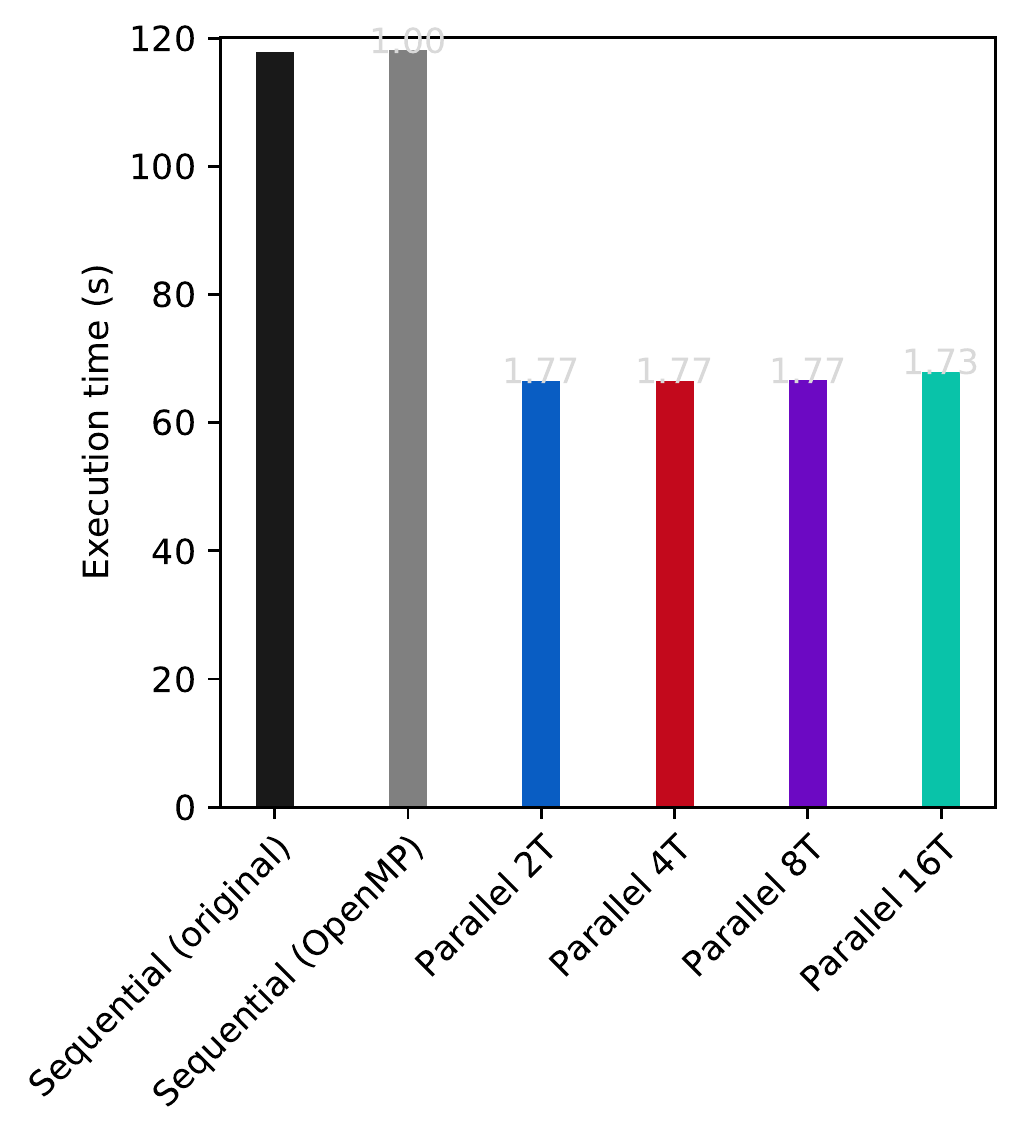}
		\caption{Molecular-Dyn WS}
		\label{fig:res-molecular-dyn-ws}
	\end{subfigure}%

	\caption{Execution details for the two test cases on PC and WS configurations.
		The speedup against the sequential execution is given above each bar.}
	\label{fig:res-pc-ws}
\end{figure}

Figure~\ref{fig:res-pc-ws} provides the results for both configurations and test cases.
The execution time is given for the original version, the parallel version executed in sequential and the parallel version executed in parallel.
We observe that for the Quicksort, Figures~\ref{fig:res-quicksort-pc} and~\ref{fig:res-quicksort-ws}, the parallel version executed in sequential suffers from a small overhead.
However, this overhead vanishes for the Molecular-Dyn application, Figures~\ref{fig:res-molecular-dyn-pc} and~\ref{fig:res-molecular-dyn-ws}.

Concerning the parallel executions, APAC provides a speedup for the Quicksort up
to  4  threads  on  the  PC  configuration  and up  to  16  threads  on  the  WS
configuration.  For  the Molecular-Dyn  application, the parallel  efficiency is
quickly bounded and there  is no improvement with more than 2  threads on the PC
and with  more than 4 threads  on the WS.   This limitation comes from  the fact
that we measure the elapsed time of  the whole execution and because there is an
overhead  that remains  significant due  to the  code's taskification  at places
where execution will finally be sequential.

However, the results are promising because we can parallelize codes that were not designed to be executed in parallel and to provide a significant speedup compared to the sequential execution.
\vspace{-0.5cm}
\vspace{-0.5cm}
\addtocounter{figure}{-1}
\section{Conclusion}
\label{sec:conclusion}
We  have presented  the  APAC compiler  conceived  to automatically  parallelize
sequential programs.  We have described  the different mechanisms implemented in
APAC that allow us to extract  data dependencies and to produce OpenMP compliant
source code.  This  includes mechanisms for managing the number  of tasks or the
depth of parallelization  manually.  We evaluated our approach  on two classical
computational  problems,  and  demonstrated  that  APAC  enables  a  significant
speedup.
\section{Discussion and Perspective}
\label{sec:discussion}
The path to creating a fully  automatic and efficient task-based parallel system
will require  leveraging several challenges  that we plan to  address.  Firstly,
the limitation of  the number of tasks  is currently manual in APAC  and must be
set by  the user, which  requires an understanding of  how the code  is actually
parallelized.  Therefore, the  final system must provide  an automatic strategy.
Secondly, we  should evaluate the  granularity of  tasks statically or  build an
incomplete granularity model  that would be completed at runtime  to decide when
to enable  the tasks.   We should  also aggregate tasks  that are  contiguous if
there are  dependencies between them,  aggregate tasks  if the granularity  of a
single task seems inefficient or avoid creating  tasks if there is only one task
in a \emph{taskgroup} section.  Thirdly, we  must detect range accesses but this
cannot be done by  considering each value in the range  as a distinct dependency
element.  If  a function works  on an array of  millions of elements,  we cannot
declare millions  of dependencies  for the task.   That is why  we will  need to
investigate if the dependency interval should  be used or if macro-dependency is
sufficient.  Fourthly, we  should transform portions of code that  are not calls
into  tasks.  Finally,  we  should detect  potential cases  that  could lead  to
speculative execution.
\clearpage
\bibliography{apac2020}
\clearpage
\section*{Appendix}
\begin{figure}[H]
\centering
\vspace{-1cm}
	\begin{subfigure}[c]{\textwidth}
	\begin{lstlisting}[escapechar=|]
void a_function(const int a, const int* b, int& c){
    /* This function does not perform any call */
    /* Therefore, no taskgroup is added to it */
    ... Some code ...
}
void a_function_with_call(const int a, const int* b, int& c){
    /* This function has one call */
    /* So, a taskgroup contains its body */
    #pragma omp taskgroup
    {
        /* Each call becomes a task */
        #pragma omp task depend(in:a,b) depend(inout:c) default(shared) |\label{codex:TransformedBase:task2}|
        {
            a_function(a,b,c);
        }
    }
}
int main(){
    /* The parallel region is declared in the main */
    #pragma omp parallel |\label{codex:TransformedBase:parallel}|
    #pragma omp master
    #pragma omp taskgroup
    {
        int a; int *b; int c;
        /* Each call becomes a task */
        #pragma omp task depend(in:a,b) depend(inout:c) default(shared) |\label{codex:TransformedBase:task1}|
        {
            a_function_with_call(a,b,c);
        }
    }
    return 0;
}
\end{lstlisting}
\caption{Automatic parallelization code.}
\label{code:TransformedBase}
\end{subfigure}
\captionof{Code}{Example of transformation - automatic parallelization.
               The compiler first creates the parallel region in the \emph{main} followed by a \emph{taskgroup}.
               Then, each call is put into a task, and the arguments are marked as \emph{in} or \emph{inout}.
               All the OpenMP statements of this example have been added by APAC.}
\end{figure}
\begin{figure}[H]
	\centering
		\begin{subfigure}[c]{\textwidth}
		\begin{lstlisting}[escapechar=|]
int x;
const int y = f(x);
\end{lstlisting}
\caption{Original code.}
\label{code:OriginalAssignment}
\vspace{0.5cm}
		\end{subfigure}
		\begin{subfigure}[c]{\textwidth}
		\begin{lstlisting}[escapechar=|]
int x;
int y;
#pragma omp task depend(in:x) depend(inout:y) default(shared)|\label{codex:assignment}|
{
    y = f(x);
}
\end{lstlisting}
\caption{Transformed code.}
\label{code:TransformedAssignment}
		\end{subfigure}
	\captionof{Code}{Example of transformation - Assignment.
	                 The compiler removes the \emph{const} qualifier from \emph{y}, then it declares the \emph{y} before the task, and finally creates a task where the result from \emph{f(x)} is put into \emph{y}.}
\end{figure}
\begin{figure}[H]
\centering
\vspace{1cm}
        \begin{subfigure}[c]{\textwidth}
		\begin{lstlisting}[escapechar=|]
if(...){
    int var1;
    int& var2 = z;
    ... some code ...
}
\end{lstlisting}
\caption{Original code.}
\label{code:OriginalIfDecl}
\vspace{0.5cm}
		\end{subfigure}
		\begin{subfigure}[c]{\textwidth}
		\begin{lstlisting}[escapechar=|]
if(...){
    int* apac_ptr_var1 = new int(); |\label{codex:newptr}|
    int& var1 = *apac_ptr_var1;
    int& var2 = z;
    ... some code with tasks...
    #pragma omp task depend(inout: var1) firstprivate(apac_ptr_var1) default(shared)
    {
        delete apac_ptr_var1;
    }
}
\end{lstlisting}
\caption{Transformed code.}
\label{code:TransformedIfDecl}
		\end{subfigure}
	\captionof{Code}{Example of transformation - Declaration in scope.
	                 The compiler allocates \emph{apac\_ptr\_var1} in the heap and transforms \emph{var1} as a reference on it.
	                 The main code remains unchanged and continues to use \emph{var1}.
	                 Finally, \emph{apac\_ptr\_var1} is deleted by an extra task at the end of the scope.}
\end{figure}		
\begin{figure}[H]
\centering		
		\begin{subfigure}[c]{\textwidth}
		\begin{lstlisting}[escapechar=|]
int var;
f(var);
... some code ...
var += 1;
\end{lstlisting}
\caption{Original code.}
\label{code:OriginalSynchronization}
\vspace{0.5cm}
\end{subfigure}
		\begin{subfigure}[c]{\textwidth}
		\begin{lstlisting}[escapechar=|]
int var;|\label{codex:synchro}|
#pragma omp task depend(inout: var) default(shared)
{
    f(var);
}
... some code with tasks ...
#pragma omp taskwait
var += 1;
\end{lstlisting}
\caption{Transformed code.}
\label{code:TransformedSynchronization}
		\end{subfigure}
                \captionof{Code}{Example    of   transformation    -   Coherency
                  enforcement.  If a variable is used  in a task and in the body
                  of     the     function,     the    compiler     inserts     a
                  synchronization/\emph{taskwait} to make sure  that the task
                  has completed.}
\end{figure}
\clearpage
\begin{figure}[ht!]
\centering		
		\begin{subfigure}[c]{\textwidth}
		\begin{lstlisting}[escapechar=|]
int functionWithReturn(int a, int b){
    if(a>b){
        ... some work with a ...
        return a;
    }
    else{
        ... some work with b ...
        return b;
    }
}
\end{lstlisting}
\caption{Original code.}
\label{code:OriginalReturn}
\vspace{0.5cm}
		\end{subfigure}
		\begin{subfigure}[c]{\textwidth}
		\begin{lstlisting}[escapechar=|]
int functionWithReturn(int a, int b){|\label{codex:return}|
int apac_res;
#pragma omp taskgroup
{
    if(a>b){
        ... some work with a ...
        #pragma omp taskwait
        apac_res = a;
        goto apac_endtaskgrouplabel_functionWithReturn;
    }
    else{
        ... some work with b ...
        #pragma omp taskwait
        apac_res = b;
        goto apac_endtaskgrouplabel_functionWithReturn;
    }
apac_endtaskgrouplabel_functionWithReturn: ;
}
return apac_res;
}
\end{lstlisting}
\caption{Transformed code.}
\label{code:TransformedReturn}
		\end{subfigure}
	\captionof{Code}{Example of transformation - Return management.
	                 Each return statement is replaced by a \emph{taskwait} followed by an assignment to \emph{apac\_res} and finally by a \emph{goto} to jump at the end of the \emph{taskgroup}.
	                 Finally, \emph{apac\_res} is returned after the \emph{taskgroup}.}
\end{figure}
\end{document}